# Topologically Linked Crystals

Toru Matsuura [a], Masanori Yamanaka [b], Noriyuki Hatakenaka [c], Toyoki Matsuyama [d], and Satoshi Tanda [a, *]

[a] Department of Applied Physics, Hokkaido University, Sapporo 060-8628, Japan
[b] Department of Physics, College of Science and Technology, Nihon University, Tokyo 101-8308, Japan
[c] Faculty of Integrated Arts and Sciences, Hiroshima University, Higashi-Hiroshima, 739-8521, Japan
[d] Department of Physics, Nara University of Education, Nara 630-8528, Japan

**Abstract**

We discovered a new class of topological crystals, namely linked rings of crystals. Two rings of tantalum triselenide ($TaSe_3$) single crystals were linked to each other while crystal growing. The topology of the crystal form is called a "Hopf link", which is the simplest link involving just two component unknots linked together exactly once. The feature of the crystals is not covered by the conventional crystallography.

## 1. Introduction

Exotic topological crystals, such as ring-shaped crystals, Möbius strips of crystals, and figure-of-eight ($2\pi$-twisted strip) crystals, have been successfully created in niobium (Nb) and selenium (Se) compounds ($NbSe_3$) [1, 2], despite their inherent crystal rigidity. These crystals have topological forms consisting of closed strips with *one component* [3]. And they have opened a new field of investigation related to topological effects in quantum mechanics and nanotechnology [4-11].

Furthermore, we discovered new topological crystals in tantalum (Ta) and selenium (Se) compounds ($TaSe_3$), which are two ring-shaped crystals linked to each other exactly at once. Due to the link the two rings cannot be separated without cutting of chemical bonding. The linked rings of crystals are categorized in a higher class of topological crystals because the topology involves *two components*. This topology is called a "Hopf link", which is the simplest link involving just two component unknots [3]. Hopf links are realized in soft materials such as molecules [12] and DNA [13], which are known as "catenanes." However, it is very surprising that rigid crystals have these topological forms. The discovery of Hopf link of crystals introduces the concept of framed knots [3] in crystallography. In this paper, we report the growing condition of the crystals and propose a model of formation mechanism of links.

## 2. Crystal features

Ring-shaped crystals of $TaSe_3$ have been discovered in 1999 [14], following the discovery in $NbSe_3$. $TaSe_3$ crystals are known to be a one-dimensional inorganic conductor that exhibits superconductivity below 2 Kelvin [15, 16]. $TaSe_3$ belongs to the same crystalline family as $NbSe_3$. These crystals are needle-like crystals called whiskers, since they grow extremely faster along b-axis direction than along a-axis and c-axis directions due to their one-dimensional crystal structures. Therefore they can be bent or twisted more easily than three-dimensional crystals, resulting in the coexistence of local crystal structures and global topological forms [1]. As a result, these crystals possess similar characteristics to soft materials.

Figure 1 shows a scanning microscope image of linked crystal rings with diameters of 54 and 46 μm. Both rings are single crystals of TaSe$_3$. The red and blue ring-shaped crystals are linked exactly once. The linked rings of crystals have unusual properties associated with framed knots. By several approaches such as X-ray diffraction, electron-beam diffraction, and electric transport measurement, it was confirmed that b-axis direction of ring-shaped crystals of TaSe$_3$ and NbSe$_3$ head along their circumference direction [17]. Therefore, the linking number of the crystals can be defined by the trajectories of the b-axis. In the case of the Hopf link of crystals, the linking number is 1. Since the number is a topological invariant, uncontinuous transformation, such as cutting of chemical bond, is necessary to separate the two rings. Furthermore, if their thickness grows anyhow the two rings cannot unite into one ring-shaped crystal. These features imply that the crystallography must be expanded within the concept of framed knots.

## 3. Experimental

The linked rings of crystals have been synthesized by chemical vapor transportation (CVT) method [18, 19]. The synthesis techniques are the same as those used in previous experiments [1, 2] except that they employ an optimized non-equilibrium condition involving the circulation and supersaturation of selenium gas in a furnace.

We soaked a mixture of tantalum powder and selenium granules of 1 : 3.15 molar ratio, which was put in one of the ends of an evacuated quartz tube in an electric furnace that has a horizontal temperature gradient of 1.5 $^{\circ}$C/cm for 3 hours. The length of the tube was 20 cm and the inner diameter was 17 mm. 5 % of excess selenium was added to enhance the circulation and supersaturation of selenium gas. The highest temperature region in furnace was set at 720 $^{\circ}$C. The end of the tube with the mixture was placed at the highest temperature region. Since the temperature gradient, temperature near another end was around 700 $^{\circ}$C, namely the growth temperature of TaSe$_3$ crystals. Under the condition, the selenium gas circulation carries tantalum to lower temperature region and TaSe$_3$ crystals grow by vapor phase epitaxy at an optimal temperature region. Furthermore, the selenium gas is condensed to liquid at the lower temperature region because the gas is supersaturated. Therefore, many selenium droplets of various diameters are repeatedly created and annihilated at the lower temperature region. It is a well condition for producing topological crystals since the droplets are indispensable for producing topological crystals [1, 2].

A selenium droplet acts as a spool during crystal growth. The spooling mechanism is a fundamental process as regards crystals having topological forms. A whisker of TaSe$_3$ crystal grows on the surface of a selenium droplet by vapor and liquid phase epitaxy and encircles it along a great circle line (Figure 2A) because of its one-dimensional crystal growth. Finally its ends are connected (B), thus yielding a ring once the droplet has evaporated (C). This scenario has been verified by observation of the NbSe$_3$ crystals spooled around selenium droplets [1, 2]. Therefore, increase of number of droplets derives increase of production probability of the ring-shaped crystals and other possible topological forms in crystals.

By the optimization for producing topological crystals of TaSe$_3$, we obtained many topological crystals more than one thousand from one tube. The production ratio was about 10 times larger than that of NbSe$_3$. Almost of them were ring-shaped crystals. We found the topologically linked ring-shaped crystals in them.

## 4. Discussion

The spooling mechanism can explain the formation mechanism of the *one-component* topological crystals such as ring-shaped crystals, Möbius strip of crystals, and figure-of-eight crystals [1, 2], however, it cannot the mechanism of the link of crystals. To make a link, the whiskers on the selenium droplet must underrun another whisker at least once, or crystals must grow on a torus droplet. Nevertheless, these scenarios are physically unreasonable.

In addition to that, the Hopf link topology cannot be produced by splitting any *one component* topological crystals since the linked ring-shaped crystals were not twisted. For example, we obtain Hopf links by splitting the middle of 2π-twisted. However, this approach provides a link consisting of two 2π-twisted strips because the number of twists is conserved. Similarly, any nπ-twisted strips do not become the Hopf link of untwisted rings by splitting. Therefore, the Hopf link of crystals is categorized to new class of topological crystals.

A promising idea is that the linked rings of crystals are formed by a *successive* spooling mechanism. It seems likely that similar processes occur successively and produce linked rings of crystals. After a ring-shaped crystal is produced by the first spooling mechanism, a small selenium droplet is attached to the ring (Figure 2D), and then a new ring grows on the second droplet (E). The second ring crystal is linked once when spooling around the second droplet. Finally, linked rings of crystals are yielded after the second droplet has evaporated (F). In fact, we often observed selenium droplets attached to rings (Fig. 3) under the optimized non-equilibrium conditions. This is strong evidence for our proposed mechanism. The small droplet might be the residue of the droplet on which the first ring grew, or it might have become attached to the first ring by chance. Our proposed mechanism provides a reasonable explanation for the different diameters of the two *untwisted* seamless rings of linked crystals.

## 5. Conclusion

We have developed our experimental set-up so that it is capable of forming many droplets in a controlled manner by using an optimized non-equilibrium condition that enhances the chance of encountering a target ring. It may be possible to grow a crystal chain with a desired number of links, as well as Borromean rings [20] with *three components*, with various topologies [3] and various similar materials [21, 22] by controlling the configurations of the selenium droplets with the help of recently developed nanotechnology. Our crystals promote the expansion of the ring topology concept called framed knots [3, 24] and accelerate the reorganization of a framework of crystallography including topological molecules [20-23] unifying the notions of space group, homotopy, and embedding into curved surfaces [3].

**Acknowledgements**

This research was supported by the 21 COE program on Topological Science and Technology from the Ministry of Education, Culture, Sport, Science and Technology of Japan. The research was also supported by the Grant-in-Aid for Scientific Research (A) No. 15204029 (S. T., T. Matsuyama, and N. H.) and the Grant-in-Aid for Young Scientists (B) No.15740245 (M. Y.) of the Ministry of Education, Culture, Sports, Science, and Technology, Japan. One of the authors (T. Matsuura) is grateful to support from Research Fellowship of the Japan Society for the Promotion of Science for Young Scientists.

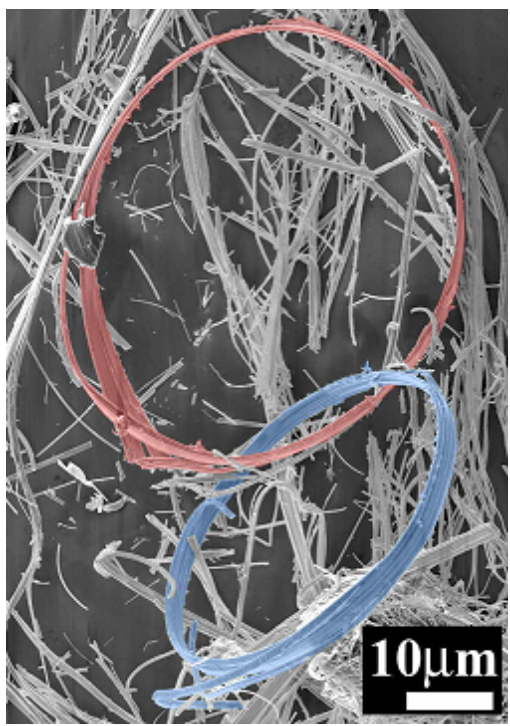

**Figure 1.** Scanning electron microscope (SEM) image of linked rings of crystal formed of a compound of tantalum and selenium (TaSe$_3$). The blue and red rings have respective diameters of 46 and 54 μm. The two rings are linked with each other exactly once.

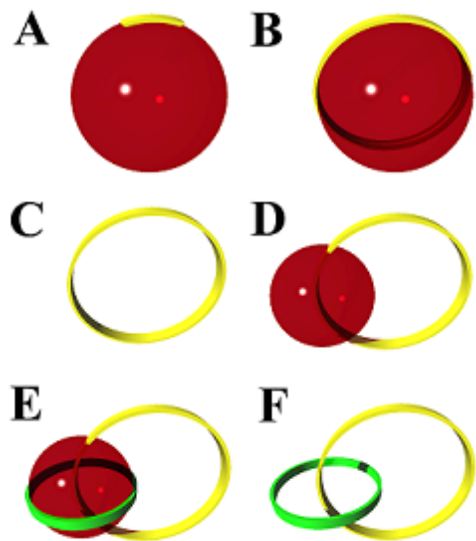

**Figure 2.** A schematic of the proposed link formation model. Under an optimized non-equilibrium condition, many selenium droplets of various diameters are repeatedly created and annihilated. Selenium droplets act as spools for thin crystals. (A) Once a thin crystal has become attached to a droplet, it grows on the droplet surface as a result of interfacial force. (B) The crystal spools around the droplets and finally the ends of the crystal are connected. (C) The non-equilibrium condition means that the selenium droplet should be annihilated leaving a ring-shaped crystal. (D) After that, a selenium droplet is sometimes attached to the ring-shaped crystal. (E) Then another ring-shaped crystal grows on the surface of the second selenium droplet. While this second crystal grows, it links with the first ring exactly once. (F) Finally, the second droplet is also annihilated leaving linked rings of crystals.

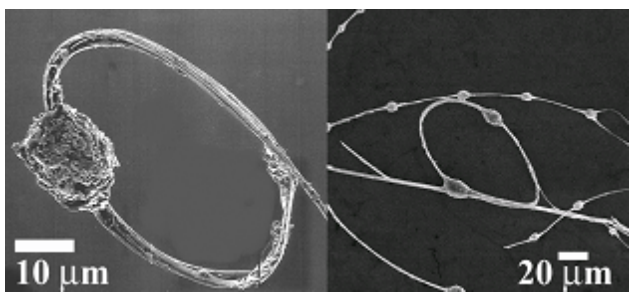

**Figure 3.** SEM images of ring-shaped crystals of TaSe$_3$ with Se condensates (droplets), which seem the starting form of the successive spooling mechanism in Fig. 2D. They are the strong evidence for the formation mechanism of linked rings of crystals proposed in this paper.